\documentclass[aps,preprint]{revtex4}%
\usepackage{amsfonts}
\usepackage{amsmath}
\usepackage{amssymb}
\usepackage{graphicx}%
\begin{document}
\title{The Local Lorentz Symmetry Violation \\and Einstein Equivalence Principle}
\author{Baocheng Zhang}
\email{zhangbc.zhang@yahoo.com}
\affiliation{School of Mathematics and Physics, China University of Geosciences, Wuhan
430074, China}

\begin{abstract}
Lorentz symmetry violation (LV) was recently proposed to be testable with a
new method, in which the effect of the violation is described as a certain
local interaction [R. Shaniv, et al, PRL 120, 103202 (2018)]. We revisit this
LV effect in the paper and show that it is not only local, but it also
represents a classical violation according to the recent quantum formulation
of the Einstein equivalence principle (EEP). Based on a harmonically trapped
spin-1/2 atomic system, we apply the results of table-top experiments testing
LV effect to estimate the corresponding violation parameter in the quantum
formulation of EEP. We find that the violation parameter is indeed very small,
as expected by the earlier theoretical estimation.

\end{abstract}
\maketitle

\section{Introduction}

Invariance under Lorentz transformations constitutes one of the fundamental
principles of modern physics. A more fundamental theory including a kind of
quantum theory for gravity, however, implies the possibility or even necessity
for Lorentz symmetry violation (LV) \cite{dm05,sl13,jdt14}. The study about
this violation has attracted extensive attention over the past twenty years,
and a nice theoretical framework called the Standard-Model Extension (SME)
\cite{cm97,cm98,kr11} was developed for the analysis of LV. This framework
formally contains all possible Lorentz-symmetry breaking terms which are
generated by the couplings between the standard fields and vacuum expectations
of the tensor fields which parameterize symmetry violations (more definitely,
it concerns only the violation of invariance under particle Lorentz
transformations, see the Ref. \cite{cm98}). The violation is always studied by
using certain terms in the presumed absence of all other symmetry breaking terms.

Although the Lorentz symmetry remains a property of the underlying fundamental
theory \cite{cm98}, it might be violated at experimentally accessible energy
scales due\ to spontaneous symmetry breaking, hence the development of
experimental tests for Lorentz invariance
\cite{kl99,kr99,mhs04,hsm09,hlb13,kv15,prp15,dfh16,vvf16}. Of particular
relevance to this aspect are several recent efforts using atoms and ions to
test LV through precision measurements. For the LV term that we are interested
in, it is usually tested by the isotropy of the speed of light (for recent
results, please refer to Ref. \cite{hsm09}). In this direction, the most
sensitive tests made use of neutral Dy atoms \cite{hlb13}, Ca$^{+}$ ions
\cite{prp15}, and Yb$^{+}$ ions \cite{dfh16}. A new method employing dynamic
decoupling was suggested to be implementable in current atomic clock
experiments \cite{sos18}, with which an especially high sensitivity to LV is
predicted. In this recent proposal \cite{sos18}, if the total angular momentum
of a physical system is fixed, the LV term can be regarded as an equivalent
term proportional to the square of the $z$-component of total angular moment
operator, a typical local interaction. In this paper, we will investigate the
related effect from such an term. Apart from revisiting this issue in the
current studies, we attempt to elucidate the meaning of locality clearly.

This interesting LV term concerns the dependence of physical results on the
moving direction and the momentum of a particle, resulting in the violation of
local Lorentz invariance (LLI) \cite{mtw70}, which also represents the
violation of Einstein equivalence principle (EEP) \cite{td12,cls15}. Since the
LV effect is always studied in those experiments made with the quantum systems
even using the obvious quantum property (i.e. entanglement in Ref.
\cite{prp15,dfh16}), we hope to investigate whether the above mentioned LV
effect is classical or quantum. This could be done by the recent suggestion
about a quantum formulation of EEP \cite{zb15,omm16}, in which the violations
of LLI can happen as either classical or quantum effects but a criteria was
given to distinguish them.

This paper is organized as follows. First, we revisit the LV term and reveal
its relation to spatial anisotropy in the second section. This is followed by
an analysis of it as a type of local interaction in the third section. In the
fourth section, we compared this LV term with the description about LLI in the
quantum formulation of EEP, and we also study the property of LV effect and
estimate the corresponding violation parameter of LLI according to the related
LV experiments. Finally, we end with a conclusion in the fifth section.

\section{Lorentz symmetry violation (LV)}

Start with the action of SME for QED \cite{vak04},
\begin{equation}
S_{SME}=\int%
\mathcal{L}%
_{SME}d^{4}x,
\end{equation}
where $%
\mathcal{L}%
_{SME}=%
\mathcal{L}%
_{\psi}+%
\mathcal{L}%
_{A}+%
\mathcal{L}%
_{g}$. $%
\mathcal{L}%
_{\psi}$ is the Lagrangian density of the fermion part and contains terms
dominating at low energies that involve ordinary matter such as protons,
neutrons and electrons and their minimal coupling with gravity and gauge
particles, $%
\mathcal{L}%
_{A}$ contains gauge fields for the photon and the fields describing particles
that are not `ordinary', like muons, mesons, neutrinos and so on and the
minimal couplings to gravity, and $%
\mathcal{L}%
_{g}$ is the gravitational part that includes the terms consisting of the
vierbein and the spin connection. When LV is considered for the electron
sector within the framework of SME, the Lagrangian density we are concerned is
the fermion part (i.e. electron) up to the first-order expansion
\cite{cm97,cm98},
\begin{equation}%
\mathcal{L}%
_{\psi}=\frac{1}{2}i\,\overline{\psi}\left(  \gamma_{\nu}+c_{\mu\nu}%
\gamma^{\mu}\right)  \overleftrightarrow{D}^{\nu}\psi-\overline{\psi}%
\,m_{e}\psi, \label{lag}%
\end{equation}
where $m_{e}$ denotes the electron mass, $\psi$ is a Dirac spinor,
$\gamma^{\mu}$ are the usual Dirac matrices and $\overline{\psi}%
\overleftrightarrow{D}^{\nu}\psi\equiv\overline{\psi}D^{\nu}\psi-\psi
D^{\upsilon}\overline{\psi}$ with $D^{\upsilon}$ the shorthand for the
covariant derivative. When the other backgrounds of spacetime like
Riemann-Cartan spacetime is considered, the vierbein and spin connection would
enter into the Eq. (\ref{lag}), i.e. see Ref. \cite{vak04} for the complete
expression. The $c_{\mu\nu}$ tensor in the above Eq. (\ref{lag}) quantifies
the strength of LV for the electron sector by the frame dependent interaction
term, which gives an energy level shift \cite{kl99,kl992,kt11},
\begin{equation}
\delta H_{\mathrm{LSV}}=-[C_{0}^{(0)}-\frac{{2U}}{{3c^{2}}}c_{00}%
]\frac{{\mathbf{p}^{2}}}{2}-\frac{C_{0}^{(2)}T_{0}^{(2)}}{6}. \label{glv}%
\end{equation}
where $\mathbf{p}$ is the momentum of the (lone) bound (valance) electron, $U$
is the Newtonian gravitational potential and the specific parameters
$C_{0}^{(0)}$ and $c_{00}$ quantifying the strength of LV have been discussed
and tested before \cite{hlb13}. It represents the violation of Lorentz
symmetry and EEP in bound electronic states. The violation of EEP within SME
frame stemmed from the dependence of effective mass on the gravitational
potential and can be tested using the transition between atomic energy levels.
In particular, the kinetic energy $\frac{{\mathbf{p}^{2}}}{2}$ would also be
changed from one energy level to another one \cite{vvf16}. But in several
other experiments, the different spin states are considered for the test of LV
\cite{prp15,sos18} and the kinetic energy remains fixed when the spin states
are changed. Thus, the first terms in Eq. (\ref{glv}) would not be observed in
these experiments. From the perspective of EEP, the first term represents the
violation of local position invariance (LPI) which usually tested by the
experiments of gravitational redshift \cite{hcm11}. The second term represents
the violation of local Lorentz symmetry.

The relativistic form of rank 2 irreducible tensor operator $T_{0}^{(2)}$ is
$T_{0}^{(2)}=c\gamma_{0}(\mathbf{\gamma p}-3\gamma_{z}p_{z})$, with $p_{z}$
the momentum component along the quantization axis fixed in the laboratory
frame. Its non-relativistic form becomes $T_{0}^{(2)}=({\mathbf{p}^{2}%
-3p_{z}^{2}})/{m_{e}}$, where $p_{z}$ is the component of the electron's
momentum along the quantization axis which is fixed in the laboratory, $m_{e}$
is the mass of the electron. Thus the second LV term $\propto C_{0}^{(2)}$ of
$\delta H_{\mathrm{LSV}}$ reduces to,
\begin{equation}
\delta H=-C_{0}^{(2)}\frac{\left(  \mathbf{p}^{2}-3p_{z}^{2}\right)  }{6m_{e}%
}, \label{hlv}%
\end{equation}
where the parameter $C_{0}^{(2)}$ characterizes the violation of Lorentz
symmetry we focus on. In relativistic physics, Lorentz symmetry implies an
equivalence of observation or observational symmetry due to special
relativity, or stated more formally that the laws of physics stay the same for
all observers moving with constant velocities with respect to one another
within an inertial frame. It has also been described sometimes as the
independence of all experimental results on the orientation or the boost
velocity of the laboratory through space \cite{mtw70}, which is seen evidently
in Eq. (\ref{hlv})

The matrix element of the $T_{0}^{(2)}$ operator is calculated using the
Wigner-Eckart theorem as \cite{prp15},%
\begin{equation}
\left\langle J,m\left\vert T_{0}^{(2)}\right\vert J,m\right\rangle
=\frac{-J(J+1)+3m^{2}}{\sqrt{(2J+3)(J+1)(2J+1)J(2J-1)}}\left\langle
J\left\vert T^{(2)}\right\vert J\right\rangle ,
\end{equation}
where $J$ and $m$ denote the quantum numbers of the total electronic angular
momentum and its projection along the quantization axis, respectively. The
term proportional to $m^{2}$ can be taken as a part of the signal for LV. If a
physical system with a fixed total angular momentum $J$ is considered, as for
the case first suggested in Ref. \cite{sos18}, the dynamics of LV term we
discuss is described equivalently by the Hamiltonian
\begin{equation}
H_{V}=\kappa J_{z}^{2},
\end{equation}
which is the very term we are concerned in this paper. In the following two
sections, we will show that this kind of interaction is local and represents
only a kind of classical violation based on the quantum formulation of EEP.

\section{Local interaction}

In earlier studies on the LV effect, different eigenstates with distinct
absolute values of the angular momentum $J_{z}$ are chosen in order to extract
the relative phase from a coherent superposition state of the eigenstates,
\textit{i.e.} in Ref. \cite{sos18}, the selected states are $\left\vert
\frac{7}{2},-\frac{7}{2}\right\rangle $ and $\left\vert \frac{7}{2},-\frac
{1}{2}\right\rangle $. The phase measurement is implemented through the
influence of the LV term on Ramsey interferometry augmented by dynamical
decoupling (DD).

The standard Ramsey interferometry consists of three parts: two $\frac{\pi}%
{2}$ pulses, and a free evolution in between. In terms of the pseudo-spin
angular momentum operators, its time evolution is described as
\begin{equation}
U_{\phi}=e^{-i\pi J_{x}/2}e^{i\phi J_{z}}e^{i\pi J_{x}/2}=e^{-i\phi J_{y}},
\end{equation}
where the two $\frac{\pi}{2}$ pulses act as 50:50 beamsplitters, and the
middle term $e^{i\phi J_{z}}$ denotes the free evolution or the free rotation
circling the z-axis (defined by a polarizing magnetic field) over an angle
$\phi$. When the LV term is present, it adds to the above free evolution and
gives%
\begin{equation}
U_{V\phi}=e^{-i\pi J_{x}/2}e^{i\phi J_{z}}e^{-i\kappa tJ_{z}^{2}}e^{i\pi
J_{x}/2}=e^{-i\phi J_{y}-i\kappa tJ_{y}^{2}},
\end{equation}
which is analogous to the nonlinear Ramsey interferometer
\cite{uf03,cb05,cs08} and provided a way for estimating the LV parameter
$\kappa$.

The effect discussed above is generally interpreted as a local (to the atom
considered) interaction. For an ensemble of $N$ atoms, the LV Hamiltonian can
be expressed as
\begin{equation}
H_{V}=\kappa J_{1z}^{2}+\kappa J_{2z}^{2}+\cdots+\kappa J_{Nz}^{2},
\end{equation}
which leads to
\begin{equation}
U_{V\phi}=\sum_{i=1}^{N}e^{-i\phi J_{yi}-i\kappa tJ_{yi}^{2}}.
\end{equation}
Evidently, this remains local, as it cannot generate any coherent results
among atoms. Consequently, such a form of local unitary interaction cannot
alter entanglement of the quantum state for the ensemble \cite{nc00}. This
also means that although the operation described by $H_{V}$ is for an ensemble
of $N$ atomic spins, the uncertainty in evaluating $\kappa$ scales as
$\propto1/\sqrt[.]{N}$, or follows the standard quantum limit (SQL). It would
be desirable to develop ideas based on entangled quantum states, such as spin
squeezed state \cite{ku93,wbi94}, the twin-fock state atomic Bose-Einstein
condensate (BEC) \cite{hb93,lpb98,lsk11,lzw17,zwl18}, the ground state of an
antiferromagnetic spin-1 atomic condensate \cite{hy00,wy16}, and so on for
improved scaling beyond the SQL.

Since the LV effect described by $H_{V}$ is local, it implies that this effect
might represent only a type of classical violation although the discussion
above is made under a quantum evolution. In the next section, we will
investigate it using a recent framework developed for testing the Einstein
equivalence principle (EEP) \cite{zb15}.

\section{Classical violation}

With the development of atomic precision measurements, some classical physical
effects that are minute and difficult to be observed, such as the violation of
weak equivalence principle (WEP), has been measured experimentally using
quantum systems. Naturally, one wants to understand whether the influence or
violation of these classical effects like WEP is actually caused by the
quantum property. To address this, Zych and Brukner \cite{zb15} constructed a
new framework to test the quantum aspects of the EEP. For a non-relativistic
quantum system with its Hamiltonian given by $H_{\mathrm{nr}}=mc^{2}%
+\frac{P^{2}}{2m}+mU\left(  Q\right)  $, where $m$ is the mass of the system,
$Q$, $P$ are position and momentum operators, respectively, for the center of
mass, and $U$ denotes the gravitational potential. In order to distinguish the
influences of different internal energies on the EEP, they suggested a quantum
formulation of the mass-energy equivalence principle by extending the mass
expression $m$ in $H_{\mathrm{nr}}$ into%
\begin{equation}
M_{k}=m_{k}I^{\mathrm{int}}+\frac{H_{k}^{\mathrm{int}}}{c^{2}},
\end{equation}
where $k=\{r,i,g\}$ represents quantities related to rest, inertial, and
gravitational masses, respectively, $I^{\mathrm{int}}$ is the identity
operator in the space of internal degrees of freedom, $H^{\mathrm{int}}$ is
the internal energy operator that can contribute to the mass, and $m$ is the
mass of the corresponding ground state for $H^{\mathrm{int}}$. Thus, the total
Hamiltonian up to the lowest order in relativistic corrections is expressed
as
\begin{equation}
H_{\mathrm{test}}=m_{r}c^{2}+H_{r}^{\mathrm{int}}+\frac{P^{2}}{2m_{i}}%
+m_{g}U\left(  Q\right)  -H_{i}^{\mathrm{int}}\frac{P^{2}}{2m_{i}^{2}c^{2}%
}+H_{g}^{\mathrm{int}}\frac{U\left(  Q\right)  }{c^{2}},\label{the}%
\end{equation}
where the validity of the EEP is guaranteed by $H_{r}^{\mathrm{int}}%
=H_{i}^{\mathrm{int}}=H_{g}^{\mathrm{int}}$ or $M_{r}=M_{i}=M_{g}$. Actually,
according to the purpose of measurements, the EEP can be divided into three
classes. One is the validity of WEP which requires the condition $M_{i}=M_{g}%
$, or $H_{i}^{\mathrm{int}}=H_{g}^{\mathrm{int}}$ if the classical WEP holds
by $m_{i}=m_{g}$. $H_{r}^{\mathrm{int}}=$ $H_{i}^{\mathrm{int}}$ indicates the
validity of LLI and $H_{r}^{\mathrm{int}}=$ $H_{g}^{\mathrm{int}}$ indicates
the validity of LPI. Therefore, if the classical WEP is not violated, the
violation from any one condition does not imply the violation of any other
one, but once any two of the three conditions is broken, the third must be
broken. Comparing Eq. (\ref{glv}) and Eq. (\ref{the}), it is seen easily that
the reasons of the violation of LPI are different. The former violation in Eq.
(\ref{glv}) stemmed from the anomalous coupling between the gravitational
potential and the external kinetic energy, and the latter violation in Eq.
(\ref{the}) is due to the anomalous coupling between the gravitational
potential and the internal energy levels. So in the future, it deserves to
make a further investigation for the possible Lorentz-breaking modifications
in the presence of gravity from the perspective of experimental observation,
since EEP is the concept of general relativity in essence. However, the
related experiments \cite{sos18} involved in this paper cannot be discussed
for the violation of LPI under the background of quantum formulation of EEP
\cite{zb15}. Further, it is noted that the violation of LLI, irrespective of
that described in Eq. (\ref{hlv}) from SME or in Eq. (\ref{the}) from quantum
formulation of EEP, is both derived from the anomalous coupling between
internal energy levels and the external kinetic energy. Thus, in order to
study the behaviour of LV using the quantum formulation of EEP, it is
convenient to assume that LPI is valid since the violation of LLI does not
imply the violation of LPI.

LLI can be tested with such a Hamiltonian \cite{zb15},%
\begin{equation}
H_{LLI}=m_{r}c^{2}+H_{r}^{\mathrm{int}}+\frac{P^{2}}{2m_{i}}-H_{i}%
^{\mathrm{int}}\frac{P^{2}}{2m_{i}^{2}c^{2}}, \label{lpv}%
\end{equation}
which constitutes an important part of the test framework for the EEP and the
violation is caused only by $H_{r}^{\mathrm{int}}\neq H_{i}^{\mathrm{int}}$ or
$M_{r}\neq M_{i}$. The terms related to the gravitational potential in Eq.
(\ref{the}) are ignored because they contribute equally to the evolution of
different energy levels (that means these terms would be subtracted when
calculating the difference of energy levels) if the LPI preserves.

After the test of the quantum aspects of EEP was suggested, some feasible
experiments were also analyzed \cite{omm16} or made \cite{rdt16}. In Ref.
\cite{omm16}, a harmonically trapped spin-$\frac{1}{2}$ atom is proposed to
test the Hamiltonian $H_{\mathrm{test}}$ and the crucial point lies in the
form%
\begin{equation}
H_{k}^{\mathrm{int}}=H_{r}^{\mathrm{int}}+\mu B\xi_{k}\left(  H_{r}%
^{\mathrm{int}}\right)  ,\text{ \ }\xi_{k}\left(  H_{r}^{\mathrm{int}}\right)
=\left(
\begin{array}
[c]{cc}%
a_{k} & b_{k}\\
b_{k}^{\ast} & c_{k}%
\end{array}
\right)  ,\text{ }\label{eps}%
\end{equation}
where $k=i,g$ represents the violation of LLI, LPI, $B$ is the external
magnetic field, and $\mu$ is the atomic magnetic moment. $H_{r}^{\mathrm{int}%
}=\frac{1}{2}\mu B\left(  \left\vert +\frac{1}{2}\right\rangle \left\langle
+\frac{1}{2}\right\vert -\left\vert -\frac{1}{2}\right\rangle \left\langle
-\frac{1}{2}\right\vert \right)  =\mu B\left(
\begin{array}
[c]{cc}%
\frac{1}{2} & 0\\
0 & -\frac{1}{2}%
\end{array}
\right)  $ and $\xi_{k}\left(  H_{r}^{\mathrm{int}}\right)  $ quantifies the
deviation from the EEP, and in particular, $\xi_{i}\left(  H_{r}%
^{\mathrm{int}}\right)  $ quantifies the deviation from the LLI, but the
required measurement was designed \cite{omm16} in the presence of gravitation
and external magnetic field. When the off-diagonal terms $b_{k}$ and its
complex conjugate are not assumed to exist, the violation of EEP parameterized
by $\xi_{k}$ is called the classical violation; otherwise, the violation is
regarded as quantum.

It is noted that $H_{k}^{\mathrm{int}}$ is actually not the test Hamiltonian,
and the complete Hamiltonian we need is obtained by inserting $H_{i}%
^{\mathrm{int}}$ into the Eq. (\ref{lpv}),%
\begin{equation}
H_{LLI}=\frac{P^{2}}{2m_{i}}+H_{r}^{\mathrm{int}}-H_{i}^{\mathrm{int}}%
\frac{P^{2}}{2m_{i}^{2}c^{2}}=\frac{P^{2}}{2m_{i}}+\mu B\left(
\begin{array}
[c]{cc}%
\frac{1}{2} & 0\\
0 & -\frac{1}{2}%
\end{array}
\right)  +\frac{P^{2}}{2m_{i}^{2}c^{2}}\left(
\begin{array}
[c]{cc}%
a_{i}^{\prime} & b_{i}^{\prime}\\
b_{i}^{\prime\ast} & c_{i}^{\prime}%
\end{array}
\right)  ,\label{lli}%
\end{equation}
where the static mass term has been ignored since it contributes only a
unrelated constant phase for the evolution of atoms. From the Eq. (\ref{lli}),
the first and second terms indicate that the atom moves in a magnetic field
freely and the $z$ direction of atomic spin is consistent with that of the
magnetic field. For example, the atom with spin-up (the eigenvalue of the
spin-$z$ component is $\frac{1}{2}$) state initially will keep in this state
during the later evolution if no extra interaction is added. The third term
represents the violation of LLI. Different from Ref. \cite{omm16}, we delete
the \textquotedblleft$\mu B$\textquotedblright\ in the third term since no
reason to imply that the assumption of LLI violation should be related to the
presence of the external magnetic field. In other words, the term
\textquotedblleft$\mu B$\textquotedblright\ is absorbed to the definition of
the violation parameters $a_{i}^{\prime}$ ($=\left(  a_{i}+\frac{1}{2}\right)
\mu B$) and $c_{i}^{\prime}$ ($=\left(  c_{i}-\frac{1}{2}\right)  \mu B$). For
the off-diagonal terms in the third term of Eq. (\ref{lli}), it will lead to
the change of the initial spin state, \textit{i.e.} the atom with spin-up
state initially will change into spin-down state (the eigenvalue of the
spin-$z$ component is $-\frac{1}{2}$). For the diagonal term, it will lead to
the shift of energy level. The corresponding shift of energy level induced by
$\xi_{i}\left(  H_{R}^{\mathrm{int}}\right)  $ has been calculated and can be
approximated as \cite{omm16}%
\begin{equation}
\Delta E_{i}^{LLI}\sim\left\langle H_{i}^{\mathrm{int}}\frac{P^{2}}{2m_{i}%
^{2}c^{2}}\right\rangle \sim\frac{\hbar\omega_{0}\left(  2n+1\right)  }%
{m_{i}c^{2}}a_{i}^{^{\prime}},\label{ess}%
\end{equation}
where the perturbation calculation gives the first order modification of the
energy level by $\left\langle n\right\vert P^{2}\left\vert n\right\rangle
=\frac{m_{i}\hbar\omega_{0}}{2}(2n+1)$, $n$ is an integer and represents the
energy level of the atom, and $\omega_{0}$ is the equivalent frequency of the
oscillator corresponding to the energy level used experimentally.

Then, we will investigate whether the LV model in the second and third section
can be put into the frame of the recent quantum formulation of EEP based on
the concrete form (\ref{lli}) constructed by the previous suggestion
\cite{omm16}. While considering the LV effect, the crucial part of the
Hamiltonian is expressed as \cite{sos18},
\begin{equation}
H_{LV}=\frac{P^{2}}{2m_{i}}+\mu BJ_{z}+\kappa J_{z}^{2}=\frac{P^{2}}{2m_{i}%
}+\mu B\left(
\begin{array}
[c]{cc}%
\frac{1}{2} & 0\\
0 & -\frac{1}{2}%
\end{array}
\right)  +\kappa\left(
\begin{array}
[c]{cc}%
\frac{1}{4} & 0\\
0 & \frac{1}{4}%
\end{array}
\right)  ,\label{lvc}%
\end{equation}
where the representation of spin-$\frac{1}{2}$ is taken to compare with the
Hamiltonian (\ref{lli}). As discussed in the last section, the LV effect was
designed to be measured using atomic Ramsey interferometry and the dynamical
decoupling method. It is easy to see from Eqs. (\ref{lli}) and (\ref{lvc})
that the above LV effect only causes the shift of energy levels but do not
cause transitions to different energy levels due to the absence of
off-diagonal terms. Hence, it only constitutes a classical violation of EEP.

Consider the example discussed before for Yb atom, and one can estimate the
parameter $a_{i}^{^{\prime}}$. As tabulated in Table I of the Ref.
\cite{sos18}, the shift of energy level caused by LV effect is%
\begin{equation}
\Delta E_{i}^{LV}\sim hC_{0}^{(2)}\times3.9\times10^{16}J\sim10^{-40}J,
\end{equation}
where $h$ is the Planck's constant and $C_{0}^{(2)}\sim10^{-23}$ according to
Ref. \cite{dfh16}. By identifing $\Delta E_{i}^{LV}$ with $\Delta E_{i}^{LLI}$
in Eq. (\ref{ess}), we obtain%
\begin{equation}
a_{i}^{^{\prime}}\sim10^{-19},\label{pv}%
\end{equation}
where the same parameters as in Ref. \cite{omm16} are adopted for the Eq.
(\ref{ess}), i.e. $n=2$, $\mu=0.429\mu_{N}$, $B=1T$, $\omega_{0}=10^{4}Hz$.
This result shows that the present experimental observation for the shift of
energy level is less sensitive than that expected in Ref. \cite{omm16} where
$a_{i}\sim1$ (the violation parameter was defined as ($a_{i}+\frac{1}{2}$)$\mu
B$ in Ref. \cite{omm16}) or $a_{i}^{\prime}\sim10^{-24}$ (according to the
definition here for the violation parameters in Eq. (\ref{lli})) is taken by
hand. A subtle point has to be clarified. For the definition of the previous
violation parameters, i.e. ($a_{i}+\frac{1}{2}$)$\mu B$, it seems that the
value of $a_{i}$ should be at the level of $1$ which is just the value taken
in Ref. \cite{omm16} to estimate the transition probability. In fact,
($a_{i}+\frac{1}{2}$) can take any value since the violation is added into the
Hamitonian by hand. So it is better to introduce a new parameter
$a_{i}^{\prime}$ to replace the previously whole expression ($a_{i}+\frac
{1}{2}$)$\mu B$ without influencing any discussion about the violation of LLI
in the quantum formulation of EEP. In this way, it is seen that the violation
parameter $a_{i}$ that is assumed to be $1$ in Ref. \cite{omm16} is equivalent
to $a_{i}^{\prime}\sim10^{-24}$. Although it is smaller than that given by Eq.
(\ref{pv}), it cannot be a constraint since the estimated transition
probability is too small to be observed at the present experimental presicion
to the best of our knowledge. Therefore, we can declare that the Eq.
(\ref{pv}) represents a constraint for the violation parameter from the
present experiments \cite{dfh16,sos18}, unless a better constraint is found
from any other experiments.

It is noted that when we constrain the violation parameter $a_{i}^{\prime}$
with the atomic experiments related to the LV test (it is easy to repeat the
above related discussion to give the constraint for the violation parameter
$c_{i}^{\prime}$ $\sim10^{-19}$), we have assumed that the discussion about
the LV is based on one specific model for LLI violation in the quantum
formulation of EEP, in which the specific model is given by the atom of
spin-$\frac{1}{2}$ state placing in an external magnetic field. Whether the
phenomenological description for the quantum formulation of EEP is consistent
with SME theory in essence required a further investigation.

\section{Conclusion}

In this paper, we revisit the LV effect and its influence on the Ramsey
interferometry. We provide a clear expression for the atomic Ramsey signal in
the absence or presence of the LV effect. It is emphasized that this type of
the LV effect discussed presents itself only as a local influence which limits
its detectability using the experiments of quantum entanglement beyond the
standard quantum limit unless a new understanding is found or suggested. The
LV effect can be regarded as equivalent to a Hamiltonian proportional to the
square of the angular momentum component along the z-axis, which can be
compared with the recently suggested quantum formulation of the EEP. We
compare the LV effect and the description for the LLI in the quantum
formulation of EEP, and find that the LV effect only amounts to a kind of
classical effect which cannot generate the non-local influence unitarily to
change entanglement among atoms. Based on the spin-$\frac{1}{2}$ system, we
build the relation to constrain the classical violation within the new EEP
test framework by the precision experiments performed to test the LV. Although
the built relation requires the further confirmation, in particular in the
presence of gravity, our work still provides an interesting and novel
exploration for the connection between LLI violation in the quantum
formulation of EEP and LV in SME. 

\section{Acknowledge}

This work is supported by National Natural Science Foundation of China (NSFC)
with No. 91636213 and No. 11654001. We also want to thank Li You and Lingna Wu
for their suggestions and discussions.

\end{document}